\documentclass[12pt,english]{article}
\usepackage[T1]{fontenc}
\usepackage[latin9]{inputenc}
\usepackage{geometry}
\usepackage{graphicx}
\usepackage{xcolor}
\makeatletter
\@ifundefined{showcaptionsetup}{}{%
 \PassOptionsToPackage{caption=false}{subfig}}
\usepackage{subfig}
\makeatother
\usepackage{babel}
\begin{document} 
\title{C. V. VISHVESHWARA (VISHU) ON THE BLACK HOLE TREK} 
\author{Naresh Dadhich\,,\\
Inter-University Centre for Astronomy \& Astrophysics\,, \\Post Bag 4, Pune, 411 007, India.
\\ and \\
K Rajesh Nayak\,,\\
Indian Institute of Science Education And Research Kolkata\,,\\ 
Mohanpur, Nadia - 741 246
West Bengal, India.
}
\maketitle
\abstract{With his seminal and pioneering work on the stability of the 
Schwarzschild black hole and its interaction with gravitational radiation, Vishu 
had opened a new window on black hole astrophysics. One of the interesting results that soon followed was that "a black hole has no hair", it is entirely specified by the three parameters, mass, spin and charge, and nothing more. The discovery of gravitational 
waves in 2016 produced by merger of two black holes, and observed by the 
Ligo-Virgo collaboration, carried the definitive signature of  quasi-normal 
modes, the phenomenon of black hole ringdown, exactly what Vishu had predicted in his 1970 Nature paper~(See Isaacson's commentary) 46 years ago. This was the crowning 
glory.} 

\section*{The Worldline} 

We shall begin by tracing Vishu's worldline, a brief life sketch followed by recounting, besides science, his multifaceted aspects of his life. 

C. V. Vishveshwara, who we all fondly called "Vishu", was born on 6 March 1938 in  Bengaluru. He did BSc~(1958) and MSc~(1959) from the University of Mysore. Then he 
proceeded to USA for higher studies where he first did his second Masters in Physics in 1964 
from the  Columbia University. From there, on the advice of his mentor, Robert Fuller, he proceeded to join the Charles Misner's group of gravitational physics at the University of Maryland for the doctoral studies. 
In 1967 John Wheeler coined the term black hole for the compact object described by the Schwarzschild metric, and that was also when Vishu was finishing his thesis examining its stability. He was thus the first to investigate the stability of a black hole spacetime, and that earned him a Ph.D. in 1968. 

After his Ph.D. he took up post-doctoral position at the NASA Goddard Institute for Space Studies, and moving further on the research positions at Universities of New York, Boston and Pittsburgh. In 1976, he returned to India to join Raman Research Institute where he established a school of gravitational physics. From there he moved to the Indian Institute of Astrophysics in 1992 and retired from there in 2005. In 1998, he was invited to be the founder Director of the  Jawaharlal Nehru Planetarium, Bengaluru. That is where his creative and imaginative prowess came into full bloom. The shows bear his distinctive stamp of his scholarship of mythology and culture on one hand and scientific history and beliefs on the other. The planetarium is the living tribute to his versatile genius. He was a science communicator par excellence. He remained glued to his own creation, the Planetarium, until he breathed his last on 16 Jan 2017, so much so he did review a new show to be launched just a few days before that. 

\subsection*{He had many hairs} 

Though he proscribed, black hole to have no hair, he himself had many hairs 
indicating his multifaceted interests and concerns. Besides science of which 
we would talk later in some detail, he had keen interest and appreciation of 
literature in all forms and arts in all its presentations. He was culturally accomplished person, connoisseur of good music (Indian and Western) classical as well as modern, theatre and painting. In particular, he was quite accomplished in drawing and sketching, and had a keen observation and eye of a cartoonist. His collections of cartoons laced with subtle and tickling  humour bear testimony to his skill and accomplishment in this art. 

On the other hand his book, Einstein's Enigma or Black holes in my bubble 
bath, demonstrates his prowess in prose writing of high quality and tenor. It reads like a 
travelogue novel with generous sprinkling of humour, thoughtful perceptions and  conversations. It is thoroughly absorbing and engaging, and at the same time leaves one 
thinking and wondering. This literary bug, he has inherited from his father who 
was a well known Kannada writer and scholar.   
 
\section*{Historical backdrop} 

The Schwarzschild metric was the first exact solution of the Einstein 
equations, obtained within a year of the equations were written in 1916. It  
describes gravitational field of an isolated static object. But the solution has 
unusual features at the surface, $R=2M$~(in the prevailing spirit of relativity, we would 
always set the gravitational constant and the velocity of light to unity, such that 
one measures mass in length units!) where the metric becomes singular with 
$g_{tt}=0, \, g_{rr} = \infty$. 

Then followed a long, and some times acrimonious and confusing, debate among relativists including Einstein, whether $R=2M$ represented a no real singularity. Could an astrophysical real object  be so compact to the limit that Sun's mass gets squeezed within the radius of $3$ Km? Even Einstein thought that it was not physically possible. The debate raged on unabated for nearly half a century. 

In 1939, Oppenheimer and Snyder~\cite{OpSn} , and B~Datt of Kolkata~\cite{BDatta} an year earlier, \footnote{The difference between the two studies was that latter did 
not  match the interior collapsing solution to the exterior vacuum solution 
at the  boundary. Yet it would be in fitness of things to term this as 
"Oppenheimer-Snyder-Datt" collapse. Unfortunately Datt died immediately 
afterwards and so his work was forgotten. It had now been duly acknowledged 
when it was reproduced in the Golden Oldies GRG series~\cite{BDatta} with a 
commentary.} considered gravitational collapse of homogeneous dust and 
showed that it collapsed down all the way to $R=0$ where even the Riemann curvature became  singular. This clearly indicated that a collapsing object could indeed reach the surface,  $R=2M$ where the Riemann curvature remains finite. It thus indicated that the central singularity, where the Riemann cirvature diverges, may not be unavoidable pointing to incompleteness of general relativity.    

Earlier Chandrasekhar had shown by application of quantum mechanics to equilibrium of white dwarf that if mass of an object exceeded $1.4$ times that of the Sun, electron pressure could not counterbalance gravitational pull~\cite{Chandra}, and it had to collapse further. This is the well-known Chandrasekhar mass limit for the white dwarf where electron 
degeneracy occurs. It opened up the possibility of further collapse going down 
to the neutron degeneracy, and when that happened there was nothing to check 
collapse, and that could then proceed unabated all the way down  to $R=2M$ or even beyond to the singularity, $R=0$. Thus the question of attaining such a compactness got theoretical credence and validity. The question, what does $R=2M$ physically mean, is then quite pertinent and real?  

Of course one may raise the question, homogeneous dust is a very special 
state of matter, and hence what is true for it, may not be true for a general fluid collapse. That is, of course, a valid question. Then came another remarkable work in 1953, though 
published in 1955,  from Amal Kumar Raychaudhuri of Kolkata in terms of the 
discovery of the equation bearing his name -- Raychaudhuri equation~\cite{AKR}.  
This showed inevitability of continual gravitational collapse with no reference to matter and spacetime symmetry properties, except requiring energy density plus thrice pressure should be greater than or equal to zero. Following the Raychaudhuri equation, Penrose and Hawking~\cite{P1965,HSW1965} proved the famous powerful singularity theorems -- the occurrence of central singularity is the robust prediction of general relativity.  

In 1960 Kruskal discovered \cite{KZ1960} a transformation connecting the regions $R>2M$ 
and $R<2M$, and continuously matching the two coordinate patches at the surface 
$R=2M$. This at once cleared all the confusion, and so $R=2M$ was a 
coordinate singularity caused by the bad choice of coordinates. It disappears when proper coordinates are chosen in the two patches. The so called Schwarzschild singularity thus gets demystified. It should be noted that the Riemann curvature, which measures the physical tidal force, however remained finite and regular at this surface, which was again  indicative of its spurious character.
      
About half a century later, the Schwarzschild metric was finally understood and 
realised that it  described a bizarre object, famously christened as "black hole" 
by John Wheeler in 1967. It has exotic properties that even light cannot 
escape from it, and its boundary is one way surface -- things can fall in, 
nothing can come out including light. Since no information or signal can come out, it marks a horizon for events happening inside, and so is termed as "event-horizon".  

Another remarkable  discovery arrives in 1962 in terms of the Kerr 
solution~\cite{RPK1963} describing a rotating black hole. This turned out 
even richer and more exotic than the Schwarzschild static solution in its physical properties as well as astrophysical relevance and applications. The most remarkable property of it is the dragging of space around it as it rotates; i.e. rotation is not confined to the black hole 
itself but is also shared by space surrounding it. This leads to a very interesting 
phenomenon of energy extraction from a rotating black hole, which is indeed astrophysically  very exciting. We will have something more to say about it later.

Then arrives Vishu, and the stage is set for him to explore the new bizarre object called a black hole with its very strange and interesting properties. That is what we shall take up next. 
\begin{figure}[t] 
\begin{center}
\includegraphics[scale=0.1]{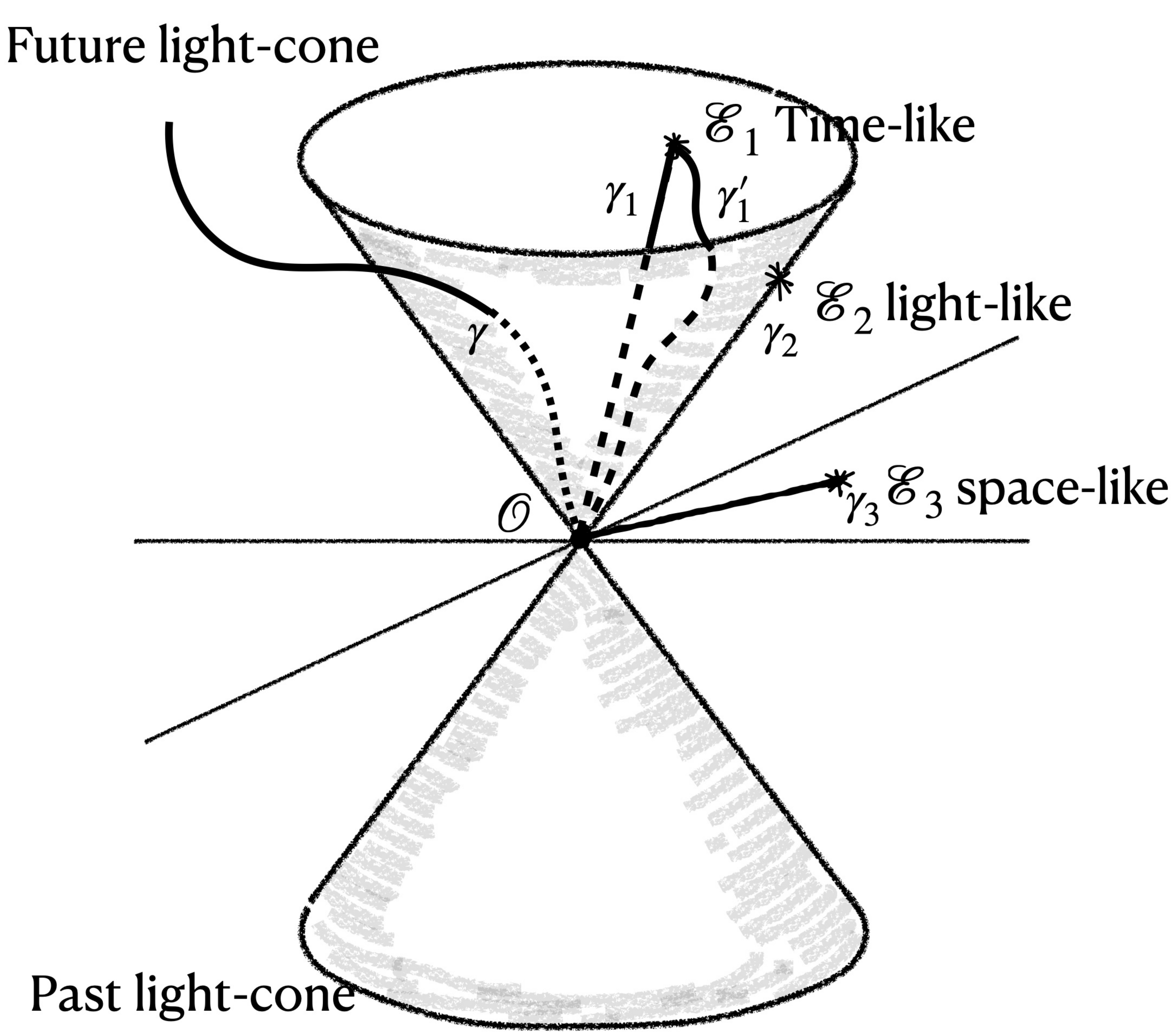}
\end{center}
\caption{Light-cone structure.  Events inside the cone can be connected to
location $\mathcal{O}$ with timelike curve.}    \label{fig:fig1} 
\end{figure}

\section*{Seminal and pioneering works} 

We shall discuss one by one the  three of Vishu's most insightful and interesting 
works, which include (1) Event-Horizon and Stationary Limit , (2) Black hole 
perturbations and Quasi-normal Modes and (3) Rotation in Black hole spacetimes .

\subsection*{Event-Horizon and Stationary Limit }

The first work involves consolidating the concept of event-horizon
in the black hole spacetimes. There are three intimately related ideas
associated with a black hole, and they are: (a) One-way membrane or event-horizon,
(b) Static or Stationary limit and (c) Infinite redshift surface.

{\bf The event-horizon:} The first concept is the idea of a one-way membrane
as the definition of event-horizon, popularly known as a black hole.
In all relativistic theories, the speed of light is the upper limit
for communication between any two points. This constraint divides
the interval between any two events in the spacetime into timelike,
lightlike or null and spacelike. This idea further develops into the
concept of the light-cone. The collection of all lightlike paths starting
from an observer's location $\mathcal{O}$ gives the light-cone, as
shown in the figure-\ref{fig:fig1}. 

\begin{figure}[t]
\begin{center}
\includegraphics[scale=0.1]{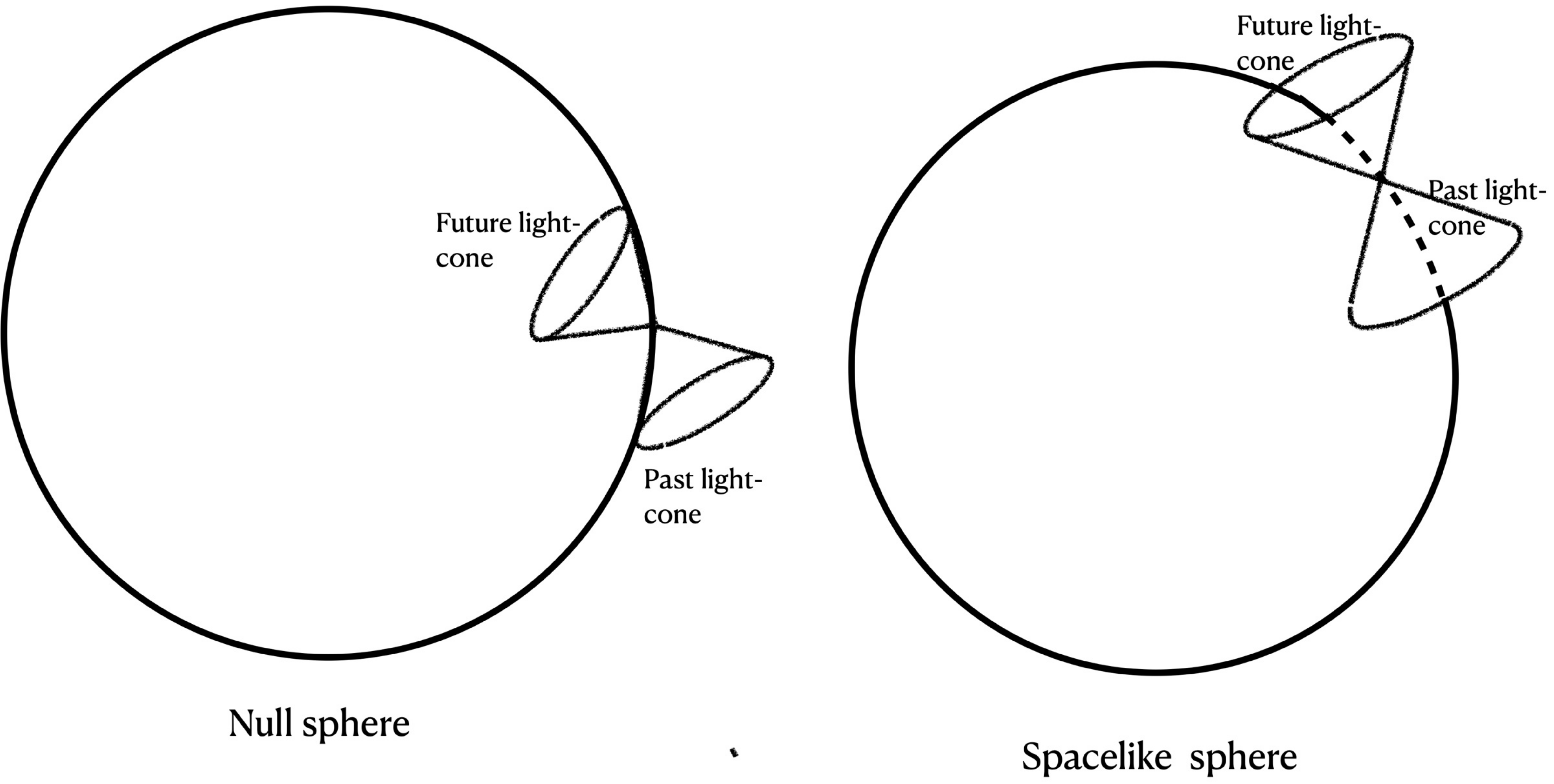}
\end{center}
\caption{Light-Cone and Event-horizon.}  \label{fig:fig2}
\end{figure}

\begin{figure}[b]
\begin{centering}
\subfloat[Timelike surface has timelike norm, ]
{\begin{centering}
\includegraphics[scale=0.06]{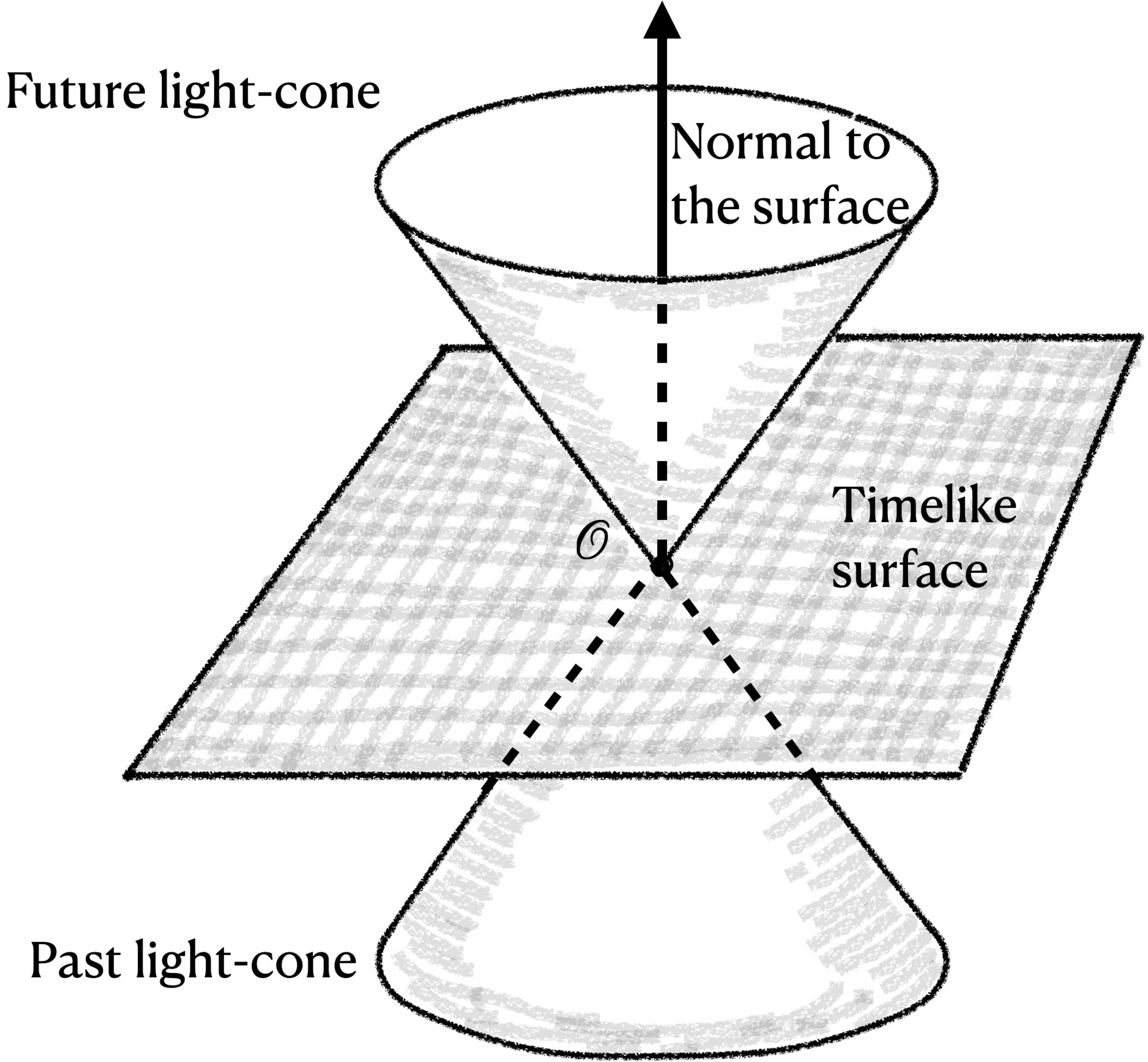}
\par\end{centering}} \hspace*{0cm}
\subfloat[Spacetime surface has spacelike  normal vector ]
{\begin{centering}
\includegraphics[scale=0.06]{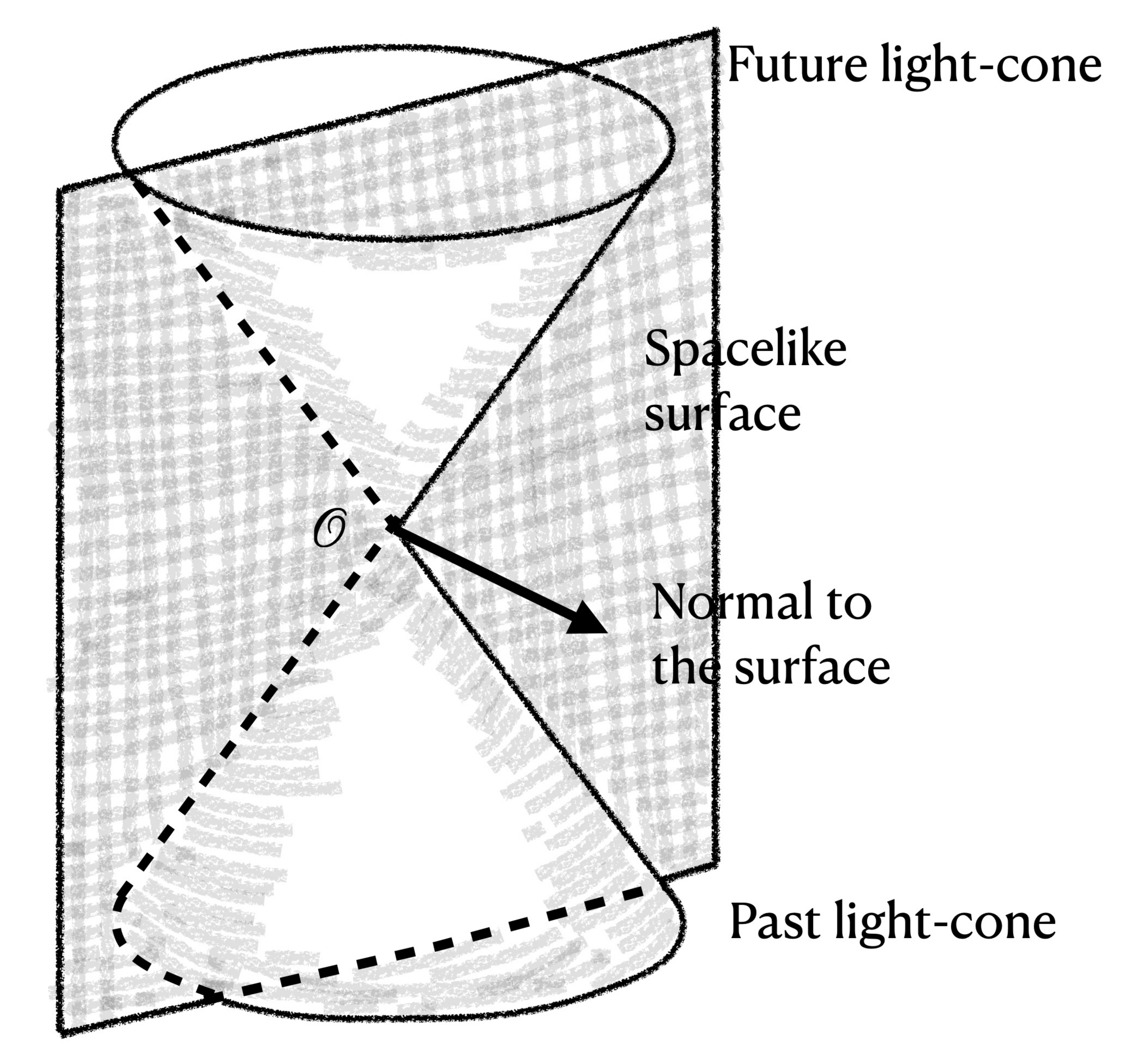}
\par\end{centering}} \hspace*{-0.0cm}
\subfloat[For null surface, norm is null vector.]
{\begin{centering}
\includegraphics[scale=0.06]{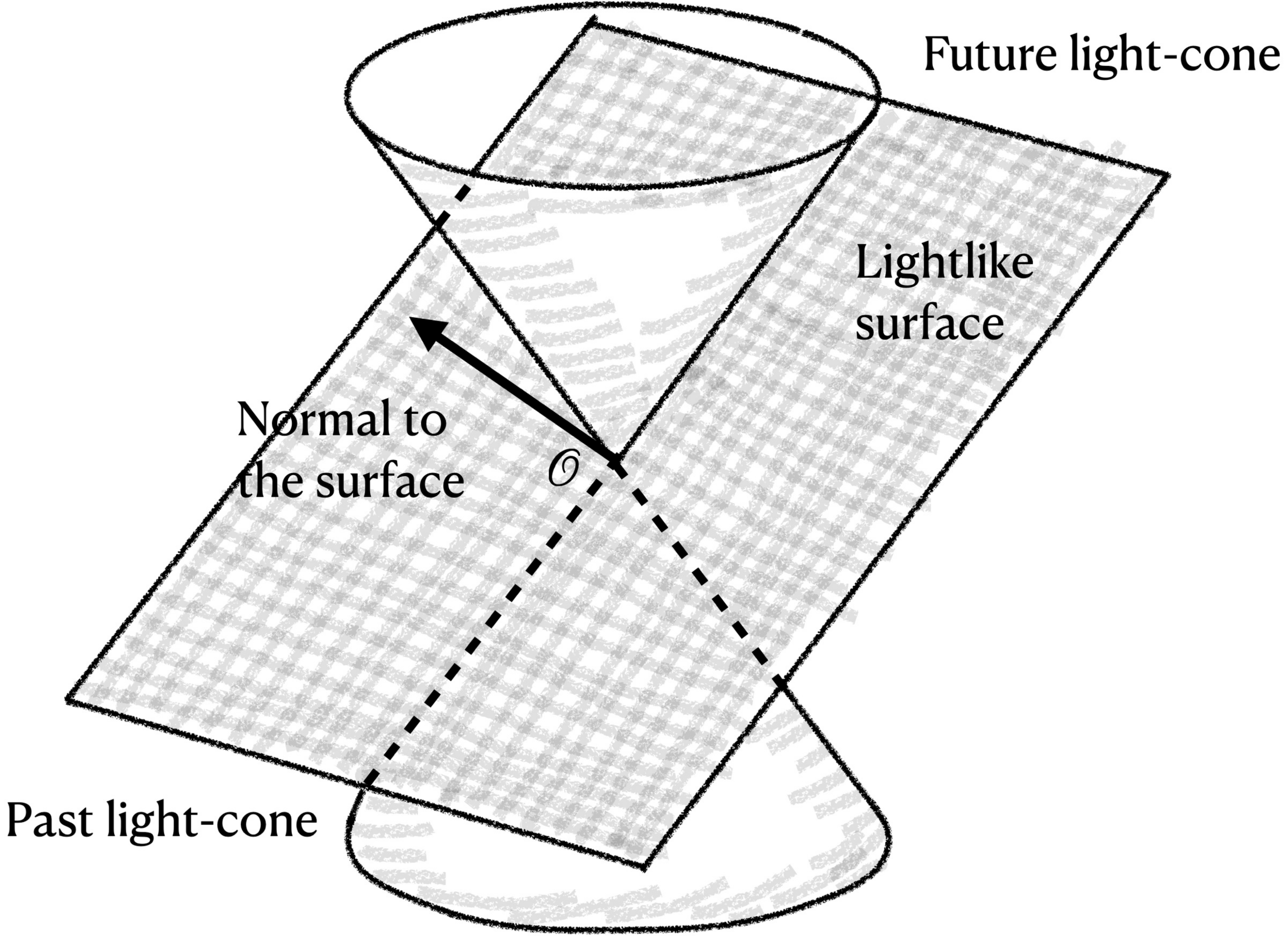}
\par\end{centering}}
\end{centering}
\caption{Timelike, lightlike and spacelike planes with light-cone}
\label{fig:fig3}
\end{figure}

A massive particle can carry a signal or message from an observer
at $\mathcal{O}$ to all the events inside the light cone. The light-cone
events are moving at the speed of light, and there is no way to reach
them from the observer at $\mathcal{O}$ without moving faster than
light. There is no way to communicate with events outside the light-cone.
A timelike curve, $\gamma$ sneaking outside light-cone can no way
come back inside again without speeding up faster than light. As one goes closer to horizon, observer's light-cone gets tilted inwards and at the horizon it is entirely pointing inwards   (Fig.2). Similarly, surfaces can be classified into timelike, null or spacelike depending
on their normal vector is timelike, null or spacelike, as shown in the figure-\ref{fig:fig3}. 
Only timelike surface having its  timelike normal can be crossed both ways, going in and out. The surfaces we generally come across, like a class room, are two way crossable -- an observer can go in and can come out. 
 
 Most fascinating of them all is the null surface. A wavefront of
light is a classic example of a null surface. The normal vector to
a wavefront is the direction vector along which light propagates,
which is null or lightlike by definition. An observer can cross a
wavefront, or a wavefront crosses an observer only once without violating
the speed limit. This crossing once is the property of all null surfaces,
and hence they are also known as one-way membranes. There is a one way surface we encounter at every epoch, $t=const.$, which we all cross only once and one way. It is different that many of us would like to cross it other way too, but can't! It is however not bounded in space. 

 The precise definition of a black hole is an inward or future pointing null surface that is
finite and closed. Finite and closed because an observer will not
be able to sneak out from any direction. The direction is along inward
such that an observer would be able to go in but not come out, as
shown in the figure-\ref{fig:fig2}. 

However there is a common sense indicator of horizon where a freely falling massive particle attains the velocity of light. That is, as in the Newtonian gravity, velocity is given by $v^2 = 2\Phi(R)$ where $v=\frac{\sqrt{g_{rr}}dr}{\sqrt{g_{tt}}dt}$ is the proper velocity relative a local observer and $\Phi(R) = M/R$ is the gravitational potential. This is so because the inverse square law remains unaltered in general relativity with $3$-space being curved rather than flat as is the case for the Newtonian gravity. In fact, we can say, Einstein is Newton with space curved~\cite{Dad12}.

{\bf Static limit}: The second concept associated with the
black hole is called the static limit. An observer is said to be static 
in spacetime if his/her spatial velocity is zero. When we stay at rest
on the earth's surface, the gravitational force acts downwards towards
the centre of the earth, while the floor's reaction force~(so-called
Newton's third law) acts upwards to balance the gravity. Without ground
to support, one needs to put on a rocket suit to give an upward acceleration
to remain at rest and avoid falling. Similarly, an observer can remain at 
rest around a black hole by providing an outward acceleration or rocket
suit to counter the radial pull towards the hole. These are the static observers in static spacetimes, which are spherically symmetric, such as the Schwarzschild solution (Fig. 4).

{\bf Stationary limit:} In the case of rotating black hole described by the axially symmetric Kerr solution presents a new situation. Since it is rotating about an axis, a direction gets identified and hence the spacetime has to be axially symmetric. Here a particle is subjected to pull in two different directions, one in the radial as in the static case but in addition also in the tangential direction to carry it around. This is because there is inherent rotation, indicated by the frame dragging angular velocity $\omega = -g_{t\phi}/g_{\phi\phi}$ at every point in  space surrounding the hole. That is, even a particle with zero angular momentum has to move with this angular velocity. 

It turns out that  first the angular pull becomes irresistible and that defines the stationary  limit; i.e., below this limit an observer cannot remain stay put at a location, he or she has to rotate around. That is, she or he  can though remain stationary at fixed $R$ by countering the radial pull but has to rotate around with the angular velocity $\omega$. At the stationary limit radial pull could be resisted but not the angular pull. As one goes further down when radial pull also becomes  irresistible, that is when the event-horizon is defined. For the static black hole both event-horizon and stationary limit are coincident which separate out for the rotating Kerr black hole. The region separating the two is called the ergo-region (Fig. 4). It is this that lends to rotating black hole the most exciting and interesting physical phenomena.  

The most remarkable feature of the ergo-region is that a particle can have its total energy negative relative an observer at infinity. It turns out that in this region, spin-spin interaction energy, which would be negative for counter-rotating particle, could become dominant and thereby making the total energy negative. Using this property, Penrose in 1969~\cite{Pen69} proposed an ingenious process of energy extraction (known as Penrose process) from a rotating black hole. It is envisaged that a particle of energy, $E_1$, falls from infinity and splits into two fragments in the ergo-region having energies $E_2<0$ and $E_3$. Then the fragment with negative energy, $E_2<0$, falls into the hole and the other, $E_3 = E_1 - E_2 > E_1$, comes out with enhanced energy. This is how black hole's rotational energy could be extracted out (Fig. 5). This doesn't happen for static black hole because there is no ergo-region there. It is rotation that causes the ergo-region and hence the extracted energy is rotational. 

\begin{figure}[t]
\subfloat[Static or stationary limit near black hole.]
{\begin{centering}
\includegraphics[scale=0.07]{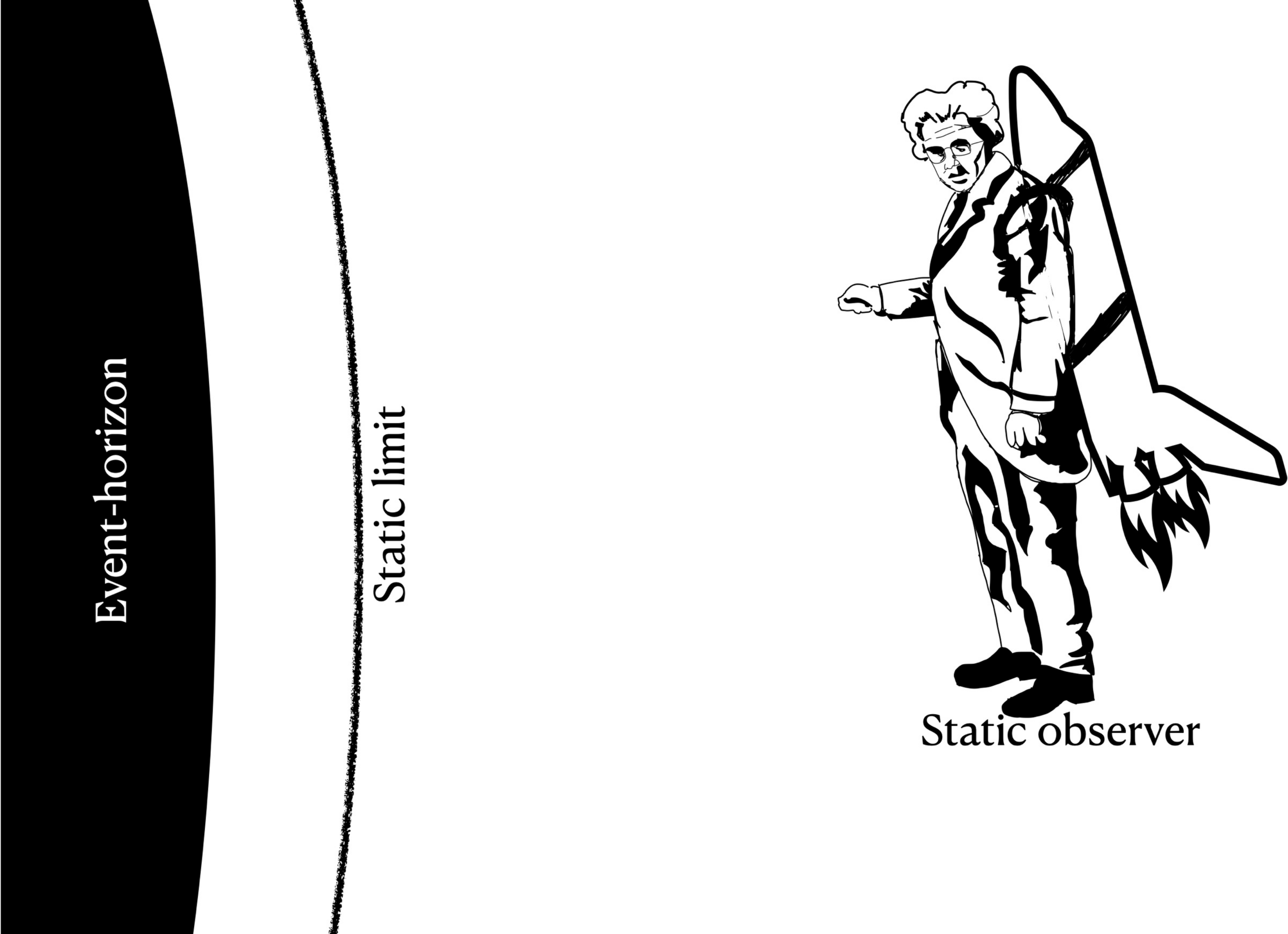}
\par\end{centering}}
\subfloat[The geometry near Kerr black hole]
{\begin{centering}
\includegraphics[scale=0.1]{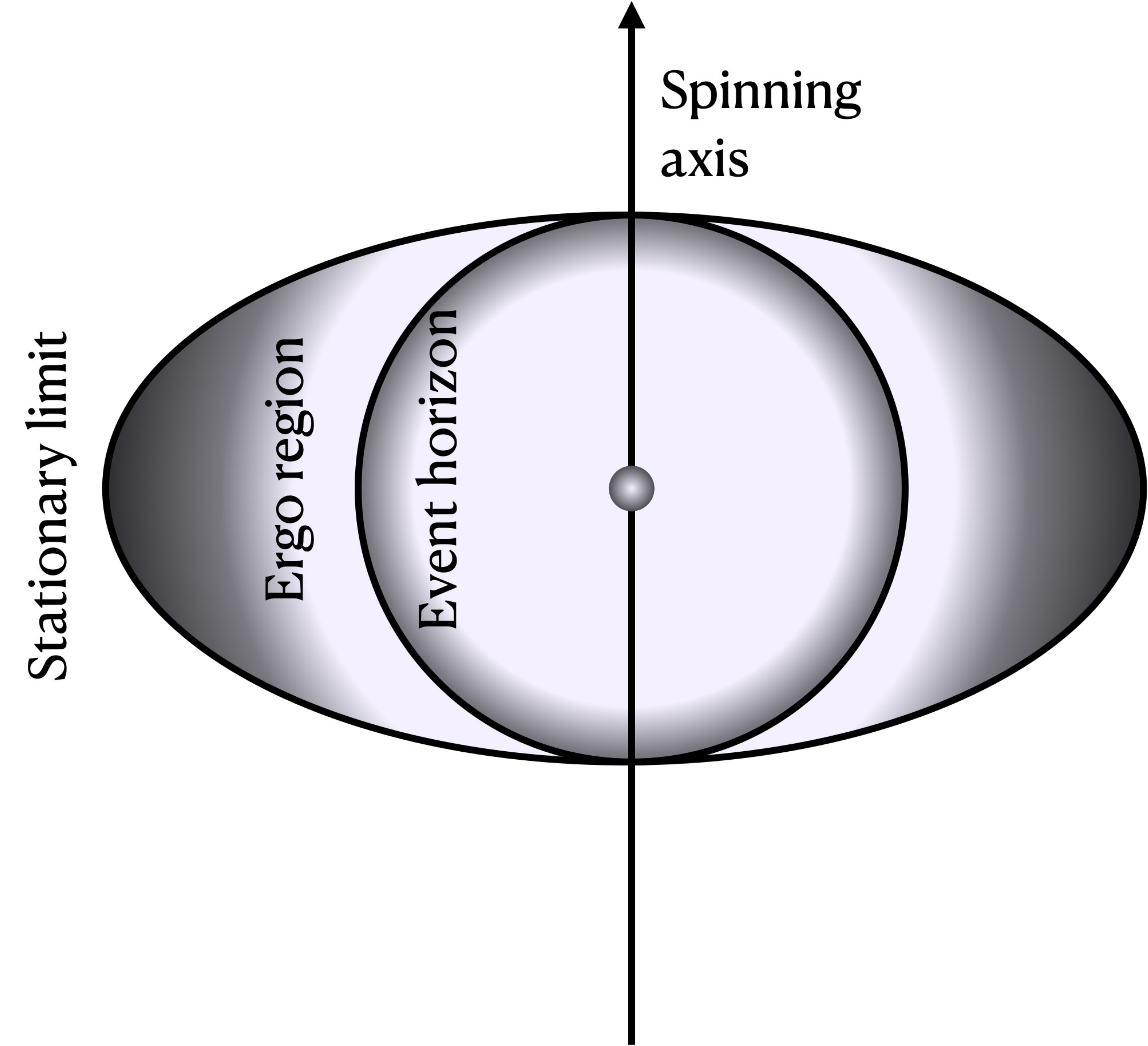}
\par\end{centering}}
\caption{}
\end{figure}

{\bf Infinite redshift surface:} The third concept is the infinite redshift
surface. When light travels from a stronger gravitational field to
a weaker gravitational field, it experiences a redshift called gravitational
redshift. As the source of light approaches the black hole, the observer
at infinity sees a more significant shift in light frequency. When
the emitter reaches $R=2M$ or $g_{tt} =0$ in general,  and light will experience infinite redshift. This limit is called the infinite redshift surface. 

In spherically symmetric black holes such as the Schwarzschild solution,
the event-horizon, static limit, and infinite redshift coincide at the 
location $r=2M$ or the Schwarzschild radius. However, rotation introduces
considerable complexity and richness of phenomena. Vishveshwara, in his work, highlighted the differences and similarity in the geometry of rotating and non-rotating
black holes. 
\begin{figure}[t]
\begin{centering}
\includegraphics[scale=0.1]{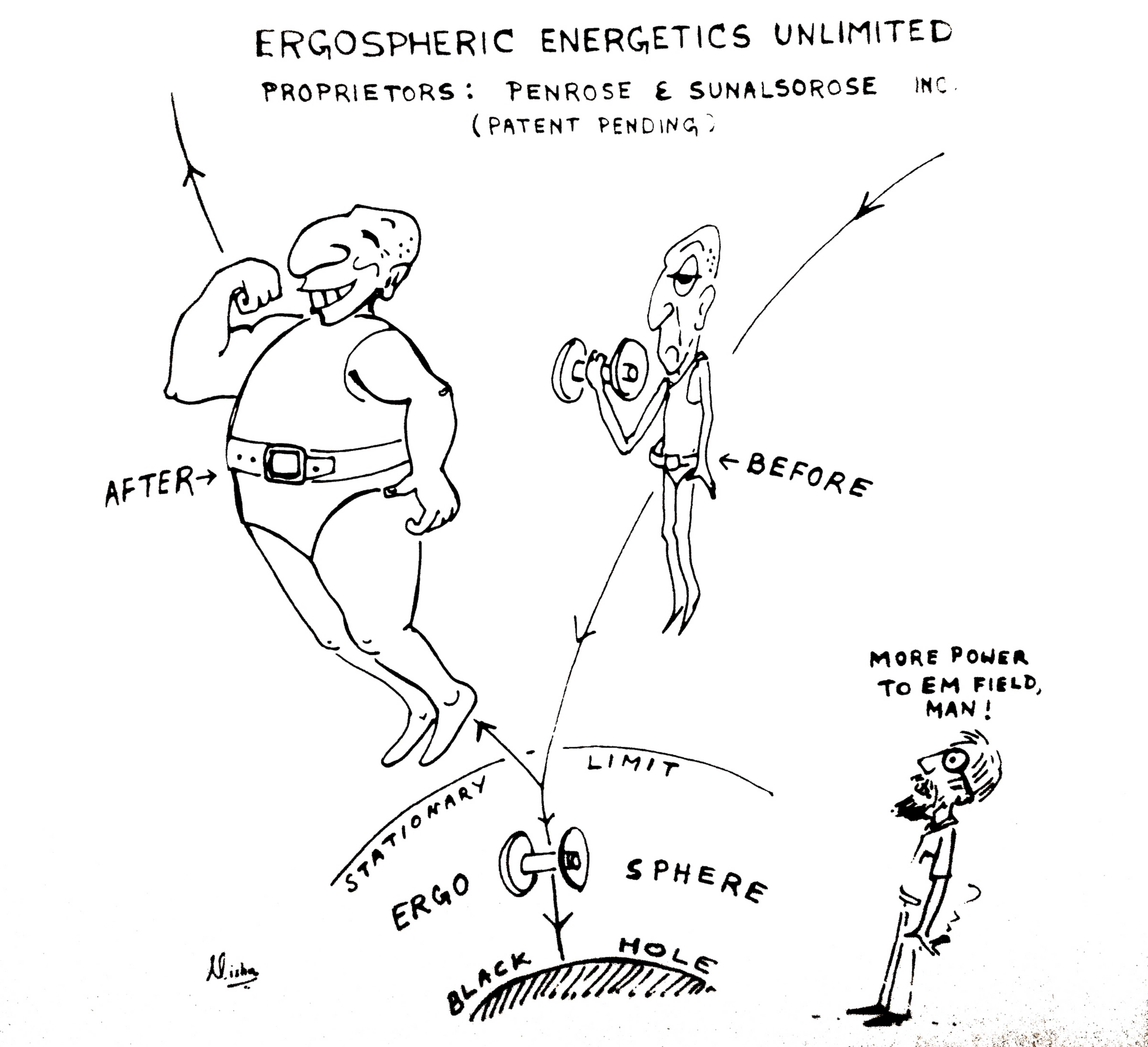}
\par\end{centering}
\caption{Ergo-region and Penrose process}
\end{figure}

\subsection*{Black hole perturbations and QNMs }

The black holes as astrophysical objects need to fulfil another important
criterion, i.e. they need to be stable under perturbations. When black holes are formed
in a stellar collapse or by the merging of two stars, often subjected
to extreme perturbations, and they need to be stable at least on the
life spend of a galaxy or the Universe itself. To examine stability
of  black hole is the problem assigned to Vishu by his supervisor
Charles Misner. Finally, it turned out to be a major topic by itself, so much so 
that Chandrasekhar had to write an over 500 pages book, The Mathematical Theory 
of Black holes, entirely devoted to black hole stability and perturbations. 

To state the problem in another way, can one destroy a black hole?
Let us try to understand the subject from an example of the ringing
of a bell. When one strikes a bell with a small hammer, a small amount
of energy is transferred to the bell and that distorts it slightly. The
bell will start ringing with notes depending on the shape and the material of the bell. 
Designed to ring for a long time, they will eventually stop ringing
when all the excess energy imparted is converted into sound wave.
Finally, it settles back to its original state. These types of perturbation
are linear, and bells are stable under linear perturbations. While
hammering, if energy transferred is considerable, the bell might get
distorted permanently or even get obliterated. 

Vishu's seminal work includes perturbing a black hole with a small
energy field. When disturbed, he first discovered that black holes
ring pretty much like a bell by emitting gravitational waves. These characteristic
modes of black holes are called Quasi-Normal Modes (QNMs) 
\begin{figure}[t]
\subfloat[Small perturbations and Quasi-normalmodes ]{\includegraphics[scale=0.09]{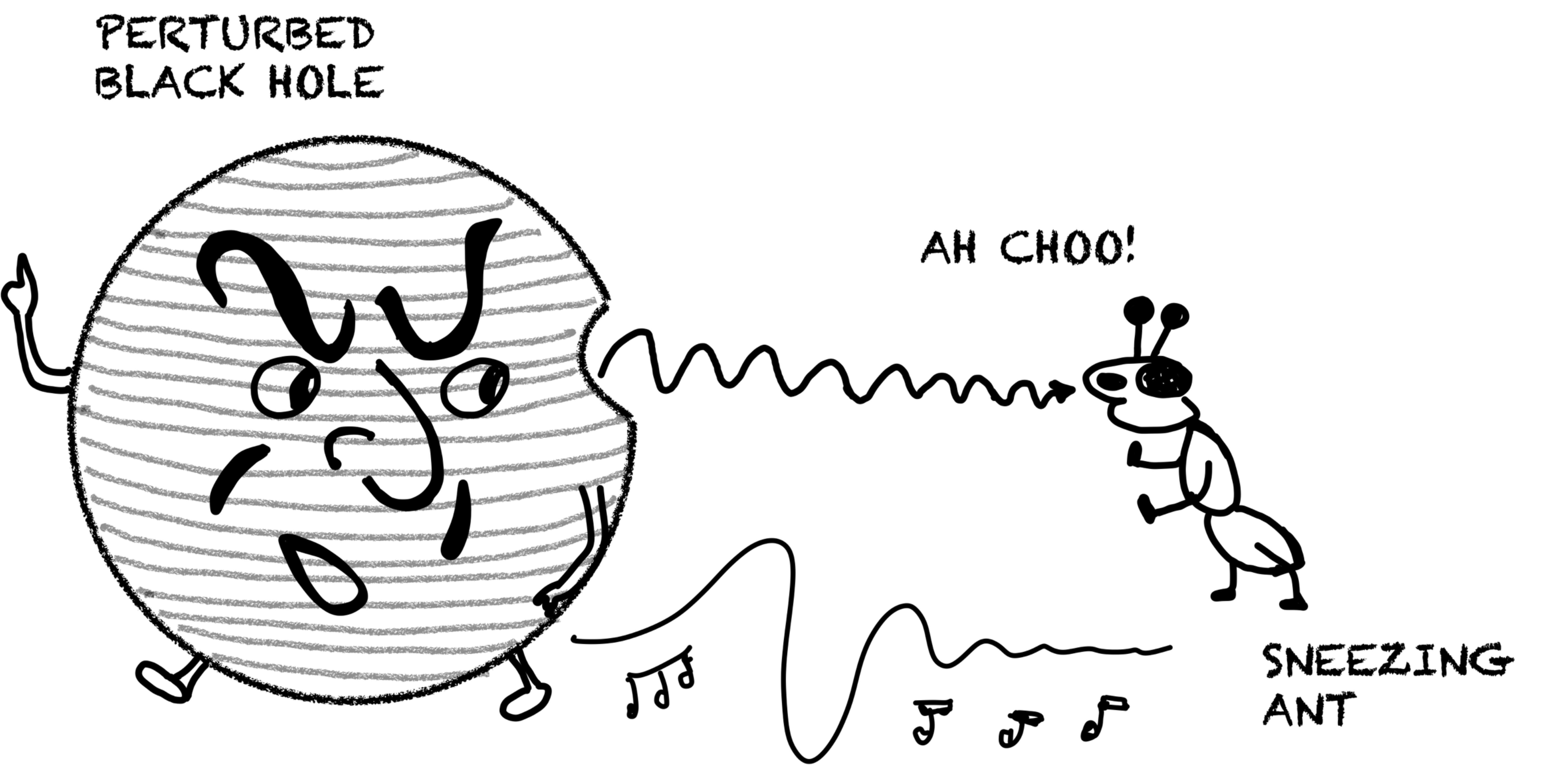}
}\hfill{}
\subfloat[Ockham razor a serious threat to the no-hair theorem]{\includegraphics[scale=0.11]{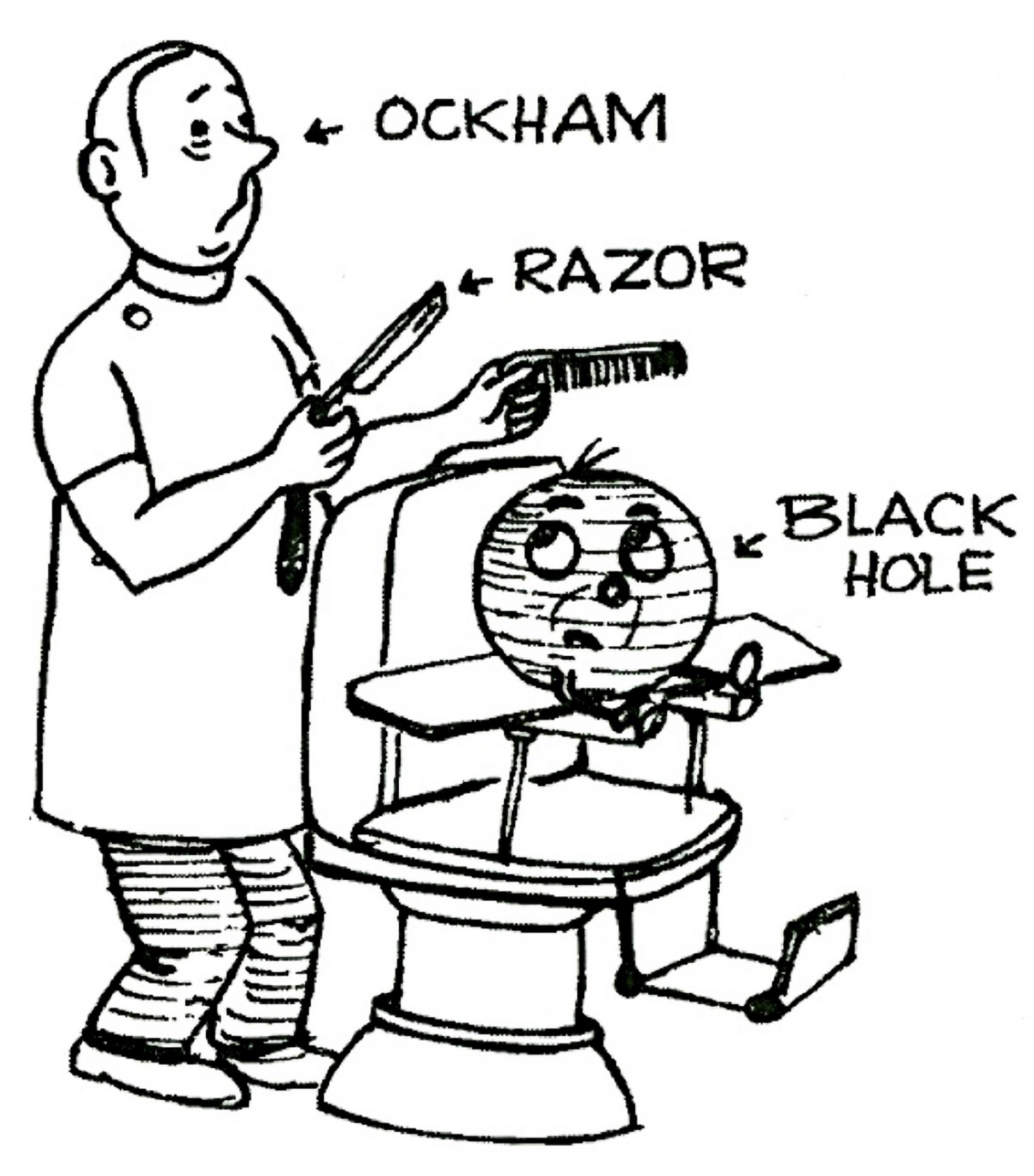}}
\caption{No-Hair theorem} \label{fig:fig5}
\end{figure}

Since a black hole can harbour only mass, charge and angular momentum, objects that fall in can only add to these three parametrs alone. All other modes get evaporated away before the object reaches the null horizon. In other words, it is a property of the closed null surface that it cannot sustain any other parameter than mass, spin and charge. This is what is popularly known as the No-Hair Theorem. 

In an amusing, but probably legendary, experiment, Galileo is said to have publicly demonstrated that two balls made up of different materials fall similarly when dropped from the leaning tower of Pisa. If the balls were to be dropped into a black hole,
they would be converted into a pile of mass and angular momentum with no trace of their material structure. So do the waves emitted by the spacetime, namely QNMs, depend only
on black hole's fundamental entities; i.e., mass, charge, and angular 
momentum [figure-\ref{fig:fig5}].

\begin{figure}
\begin{centering}
\includegraphics[scale=0.15]{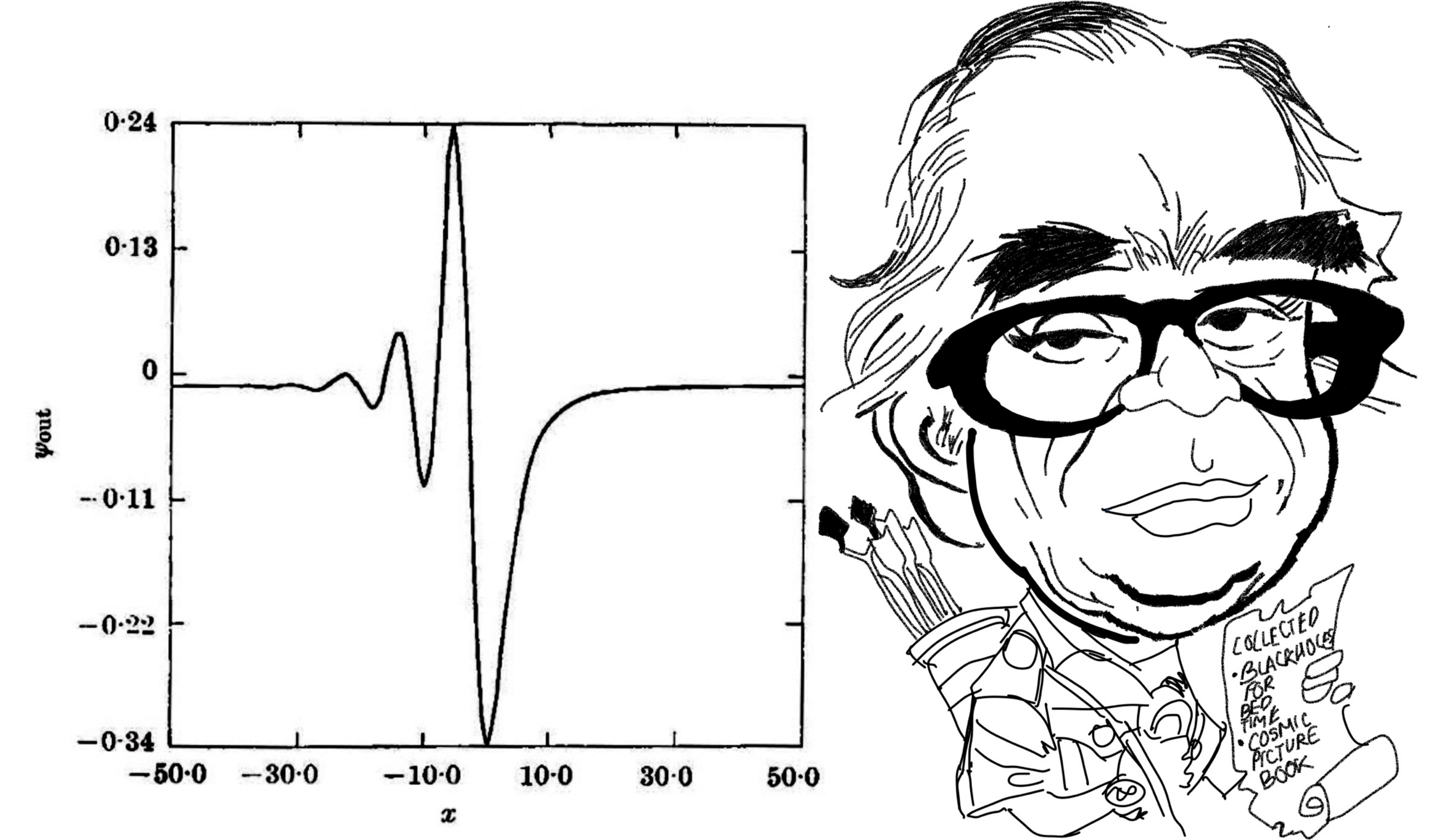}
\par\end{centering}
\caption{QNM and Vishu} \label{fig:fig6}
\end{figure}

The QNMs signals are one of the unambiguous signatures of black holes.
If detected directly, they can be a powerful tool in understanding
black hole physics. With QNMs, one may be able to distinguish between
black holes and other compact objects mimicking them. One can
verify the no-hair theorem by analysing multiple QNM modes whether they carry 
any other signature than mass, spin and charge. If any other parmeter black hole have would leave an imprint on QNMs. QNMs are the only messengers we have from black holes, the only source of information. Since existence of QNMs is prediction of general relativity, their existence hence also marks its test. [figure-\ref{fig:fig5}]. 

The first detection of gravitational waves from a black hole merger,
GW150914, by the LIGO and VIRGO collaboration is a significant step in
the direct identification of QNMs signals. The final phase of merger of black holes leaves a
highly distorted black hole, emitting QNMs. It then settles down as QNMs damp out  to a stationary state of rotating Kerr black hole. This phenomenon is called the ringdown phase.  The observed signal from GW150914 confirm the ringdown phase [figure\ref{fig:fig6}]. 
Vishu was also one of the first to use then available computing facility for analysing the QNMs as shown in the figure, and it is remarkable that the curve is pretty much alike the one in the gravitational wave discovery observation. Work is underway for identifying and characterising QNMs frequencies. For GW150914, the signal strength is weak in the ringdown phase and is not adequate enough to draw any clean inference in the context of no hair theorem. What is needed is a closer and more massive black hole merger to get a sufficiently loud signal. In the coming years, QNMs will be one of the powerful tools to probe black hole physics directly.


\section*{Rotation in black hole spacetimes }

After his work on QNMs, Vishu mainly worked on the various physical
phenomena in black hole spacetimes. Rotation brings in considerable
complexity and interesting physical effects in general, and in general relativity in particular.  We have already seen how event-horizon and stationary limit decouple in going from the Schwarzschild to the Kerr black hole. Unlike the Newtonian theory, in general
relativity, a rotating object  in addition to pulling things radially inwards, it also imparts a tangential push to take things around. That is, entire spacetime is dragged around 
it, and this is referred to as the phenomenon of frame-dragging. If it were not so, a particle very close to the horizon will have no rotational motion, as it falls into the horizon it suddenly starts rotating with black hole. There would be discontinuous jump from no rotation to rotation. If that happens, it should be reflected in some abrupt change in spacetime at the horizon which would be reflected in the Riemann curvature. Nothing abrupt happens in the curvature at the horizon. For motion to transit smoothly from outside onto the horizon,   black hole has to share its rotation with the surrounding space.  

The frame-dragging effect of the Kerr rotating spacetime is directly observable in precession of a gyroscope carried by a rest-observer staying at a fixed location. The rest-observer is experiencing a torque due to the black hole's spin -- a complex and subtle interplay between time and azimuthal angle due to rotation. It is because of this phenomenon, the rest-observers are not the best ones to probe the rotational effects of a spinning black hole. Bardeen and coworkers \cite{BRBH} proposed a locally non-rotating observers, who co-rotate with the frame-dragging angular velocity $\omega$ at the given location and remain stationary with $R=const$ but rotating. They are well defined everywhere outside the horizon. If they carry a gyroscope, it will not precess because the observer and gyroscope share the same angular velocity $\omega$. 

\begin{figure}[t]
\begin{centering}
\includegraphics[scale=0.1]{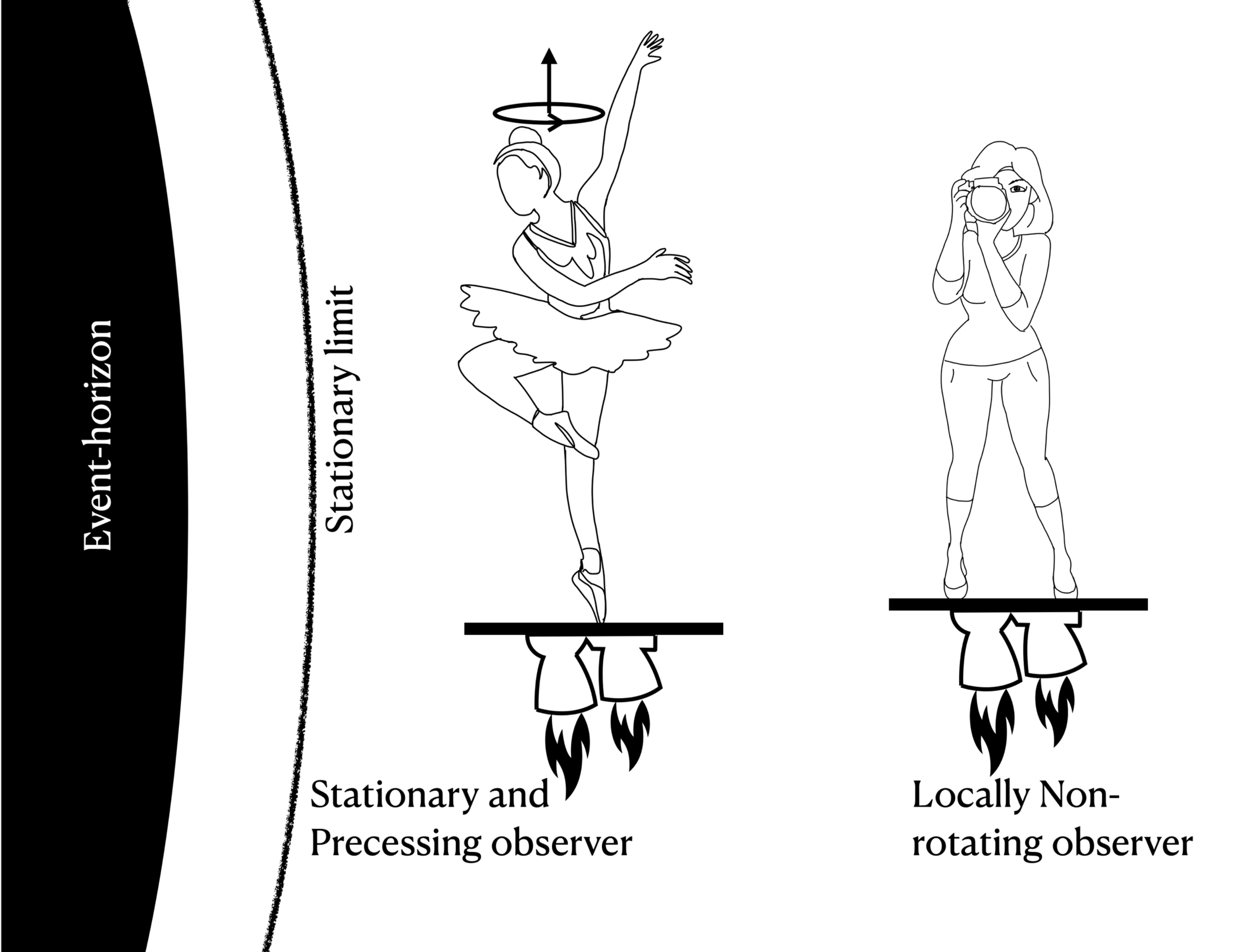}
\par\end{centering}
\caption{Stationary and Locally non-rotating observer around Kerr blackhole}
\label{fig:fig7}
\end{figure}

Vishu and his coworkers generalised the non-rotating observers to general axially symmetric
spacetimes [figure \ref{fig:fig7}]

Gyroscopes play a vital role in understanding the rotational effects
in black hole spacetimes. Precession of gyroscope can pin down the
footprints of phenomenon such as dragging of inertial frames. Gravity
Probe B indeed verified that the earth's rotational field is consistent
with  general relativity. Vishu and his long time collaborator, 
Bala Iyer, formulated gyroscopic precession \cite{IV} in the framework of Frenet-Serret
formalism. The Frenet-Serret formalism describes the geometry of curves
and is very fundamental to a set of trajectories. This formalism relates
the precession frequency of a gyroscope directly to the geometry of
the path.


\section*{ In the end }

His outstanding and impactful scientific work which provided one of the most effective and powerful tools in quasi-normal modes for black hole and gravitational wave physics as well as he set the ball rolling for probing stability of black holes, which kept the likes of Chandrasekhar engaged for nearly a decade. On the other hand, he has left a glorious legacy in science outreach and education. At the planetarium, he had initiated and promoted a number of innovative and interesting programmes such as SEED (Science Education in Early Development), SOW (Science over weekends), REAP (Research Education Advancement Programme in Physical sciences) and BASE (Bangalore Association of Science Education). 

He was man full of zeal and enthusiasm for knowing and learning everything around,  and equally keen on sharing it with others. He had a very pleasant and welcoming disposition with a great sense of subtle and tickling humour that some times turned sublime.
It was Antonio Machado, who famously said, "Traveller there is no path, Paths are made by walking", in the same vein we would like to say, 
"{\it There was no path, he made the one by walking."}  That's how we would like to remember him with affection and fondness.   
\section*{Acknowledgement}
It is a pleasure to thank Richard Isaacson for reading the manuscript and making several 
suggestions for accuracy and style. 
We wish to warmly thank Saraswati Vishveshwara for the kind permission to
use Vishu's cartoons and figures.

\end{document}